\documentclass[10pt,aps,prd,fleqn,notitlepage,superscriptaddress,nofootinbib,preprintnumbers]{revtex4-1}
\pdfoutput=1
\usepackage{amsmath,amssymb,graphicx,xspace,subfigure,relsize,todonotes}
\newcommand{\UNNLOPS}{\protect\scalebox{0.9}{UN$^2$LOPS}\xspace}
\newcommand{\UNLOPS}{\protect\scalebox{0.9}{UNLOPS}\xspace}
\newcommand{\MEPSatNLO}{\protect\scalebox{0.9}{MEPS@NLO}\xspace}

\newcommand{\POWHEG}{\protect\scalebox{0.9}{POWHEG}\xspace}
\newcommand{\MCatNLO}{\protect\scalebox{0.9}{MC@NLO}\xspace}
\newcommand{\MINLO}{\protect\scalebox{0.9}{MINLO}\xspace}
\newcommand{\Sherpa}{\protect\scalebox{0.9}{SHERPA}\xspace}
\newcommand{\BlackHat}{\protect\scalebox{0.9}{BLACKHAT}\xspace}
\newcommand{\FEWZ}{\protect\scalebox{0.9}{FEWZ}\xspace}
\newcommand{\DYNNLO}{\protect\scalebox{0.9}{DYNNLO}\xspace}
\newcommand{\abr}[1]{\langle #1\rangle}
\newcommand{\mc}[1]{\mathcal{#1}}
\newcommand{\mr}[1]{\mathrm{#1}}

\newcommand{\done}{{\rm d}}
\newcommand{\bs}{\!\!\!\!\!\!}
\usepackage[pdfborder={0 0 0}]{hyperref}
\hypersetup{
  pdfauthor={Stefan Hoeche, Ye Li, Stefan Prestel},
  pdftitle={Drell-Yan lepton pair production at NNLO QCD with parton showers},
  pdfkeywords={QCD, NNLO, Parton Shower}
}

\begin{document}

\preprint{SLAC-PUB-15961}
\preprint{DESY-14-073}
\preprint{MCNET-14-10}

\title{Drell-Yan lepton pair production at NNLO QCD with parton showers}

\author{Stefan~H{\"o}che}
\affiliation{SLAC National Accelerator Laboratory, Menlo Park, CA 94025, USA}
\author{Ye~Li}
\affiliation{SLAC National Accelerator Laboratory, Menlo Park, CA 94025, USA}
\author{Stefan Prestel}
\affiliation{Deutsches Elektronen-Synchrotron, DESY, 22603 Hamburg, Germany}

\begin{abstract}
  We present a simple approach to combine NNLO QCD calculations and 
  parton showers, based on the \UNLOPS technique. We apply the method to the 
  computation of Drell-Yan lepton-pair production at the Large Hadron Collider.
  We comment on possible improvements and intrinsic uncertainties.
\end{abstract}

\maketitle

\section{Introduction}
The combination of fully exclusive next-to-leading order (NLO) calculations in 
perturbative Quantum Chromodynamics with resummed predictions from parton showers 
has been in the focus of interest for the past decade. A wide range of matching~\cite{
  Frixione:2002ik,Nason:2004rx,*Frixione:2007vw,Hamilton:2009za,*Torrielli:2010aw,
  *Frixione:2010ra,*Alioli:2010xd,*Hoeche:2010pf,*Frederix:2011ig,*Frederix:2011ss,
  *Platzer:2011bc,Hoeche:2011fd,*Hoeche:2012ft,*Hoeche:2012fm,Nason:2012pr}
and merging~\cite{Catani:2001cc,*Lonnblad:2001iq,*Mangano:2001xp,*Krauss:2002up,
  *Hoeche:2009rj,*Hamilton:2009ne,*Lonnblad:2011xx,*Lonnblad:2012ng,Lavesson:2008ah,
  Gehrmann:2012yg,*Hoeche:2012yf,Lonnblad:2012ix,Platzer:2012bs,Frederix:2012ps} 
methods has been developed and implemented in Monte-Carlo event 
generators~\cite{Buckley:2011ms}. By now they are standard tools for simulating
final states at hadron colliders such as the LHC. However, so far only two techniques 
were implemented, that extend these methods to next-to-next-to-leading 
order (NNLO) in the strong coupling expansion~\cite{Lavesson:2008ah,Hamilton:2013fea}.
They have been used to describe two-jet production
at LEP and Higgs production via gluon fusion at hadron colliders. Event generators 
for Drell-Yan lepton pair production at NNLO QCD matched to a parton showers are not 
available. Due to the high relevance of this process as a standard candle for the LHC 
and possible future hadron colliders, we address the problem in this publication, 
and we also provide a simple formulation for matching at NNLO, improving upon 
the \UNLOPS method suggested in~\cite{Lonnblad:2012ix}.
We use the Monte-Carlo event generator \Sherpa~\cite{Gleisberg:2003xi,*Gleisberg:2008ta}, 
including a parton shower~\cite{Schumann:2007mg} based on Catani-Seymour
dipole subtraction~\cite{Catani:1996vz,*Catani:2002hc}, combined with the 
\BlackHat library~\cite{Berger:2008sj,*Berger:2009ep,*Berger:2010vm,*Berger:2010zx} 
for one-loop matrix elements. This implementation is publicly available.

Our matching scheme, which we call \UNNLOPS, preserves both the logarithmic accuracy 
of the parton shower and the fixed-order accuracy of the NNLO calculation. 
It is a generic method to augment NNLO calculations with the primitive resummation
encoded in an existing parton shower. At NLO, a considerable difference exists 
between matching methods, pertaining to the treatment of the finite remainder 
of higher-order corrections. This difference must be reduced at NNLO.
The excellent convergence of the perturbative series in the Drell-Yan process 
further reduces potential differences. We therefore expect that \UNNLOPS will
serve as a useful benchmark for future, more sophisticated NNLO matching schemes.
The parton shower employed in our calculations already includes full color 
and spin information in the first emission term and the associated Sudakov 
factor~\cite{Hoeche:2011fd,*Hoeche:2012ft,*Hoeche:2012fm}. It is therefore
improved compared to the standard large-$N_c$ approximation with spin averaging.

The outline of this paper is as follows: Section~\ref{sec:intro} gives an
introduction to the problem of matching at NLO and outlines our simplified 
approach. Section~\ref{sec:unmcatnlo} extends the simplified \UNLOPS
method to NNLO, which we dub \UNNLOPS. Section~\ref{sec:results} 
contains first results from applying the method to Drell-Yan lepton pair production 
at the LHC. We also present some benchmark results for a high-energy LHC
and a possible future proton-proton collider at 100~TeV center-of-mass energy.
Section~\ref{sec:conclusions} contains some concluding remarks.

\section{A simple example}
\label{sec:intro}
To set the stage for the discussion of our method at NNLO we reformulate
in this section the \UNLOPS method and simplify its event generation algorithm.
The extension to NNLO is then nearly straightforward. It will be presented in
Sec.~\ref{sec:unmcatnlo}.

The leading-order expression for an observable $O$ is written as
\begin{equation}
  \abr{O}^{\rm(LO)}=\int\done\Phi_0\,\mr{B}_0(\Phi_0)\,O(\Phi_0)\;,
\end{equation}
where $\Phi_0$ is the differential Born phase-space element, and $\mr{B}_0(\Phi_0)$ is
the Born differential cross section, including symmetry and flux factors
as well as parton luminosities. We now add and subtract Sudakov-reweighted 
real-emission tree-level cross sections. To the accuracy of the parton shower,
this method is equivalent to the modified subtraction in \MCatNLO~\cite{Frixione:2002ik}
and \POWHEG~\cite{Nason:2004rx,*Frixione:2007vw}
\begin{equation}\label{eq:unlops_step1}
  \int\done\Phi_0\,\mr{B}_0(\Phi_0)\,O(\Phi_0)
  -\int_{t_c}\done\Phi_1\,\Pi_0(t_1,\mu_Q^2)\,\mr{B}_1(\Phi_1)\,O(\Phi_0)
  +\int_{t_c}\done\Phi_1\,\Pi_0(t_1,\mu_Q^2)\,\mr{B}_1(\Phi_1)\,O(\Phi_1)\;.
\end{equation}
In the second term, the observable $O$ is taken at the reduced phase space point, 
determined by clustering the one-parton state $\Phi_1$ to $\Phi_0$ using
an algorithm that corresponds to inverting the parton shower~\cite{Andre:1997vh}.
We have defined the parton-shower no-branching probability for an $n$-parton state,
\begin{equation}
  \Pi_n(t,t';\Phi_n)=\exp\left\{-\int_t^{t'}\done\hat{\Phi}_1\,\mr{K_n}(\Phi_n,\hat{\Phi}_1)\right\}\;.
\end{equation}
where $\mr{K}_n$ is the sum of differential branching probabilities, including luminosity 
and flux factors for initial-state evolution as appropriate~\cite{Sjostrand:1985xi}.
The multi-particle phase space elements factorize as $\done\Phi_{n+1}=\done\Phi_n
\done\hat\Phi_1$, with $\done\hat\Phi_1$ the phase-space element for the emission
of a single additional parton. We can write $\done\hat\Phi_1=\done t\,\done z\,
\done\phi/(2\pi)J(t,z,\phi)$, where $t$ is the evolution variable of the parton shower, 
$z$ is the splitting variable, and $J$ is a Jacobian factor. $t_c$ denotes the infrared 
cutoff, and $\mu_Q^2$ is the resummation scale.
Note that the parton shower employed here covers the full resummation phase space, 
except for the region $t<t_c$. For ease of notation we have defined $t_1=t(\Phi_1)$.

Equation~\eqref{eq:unlops_step1} describes the one-parton state in the simplest possible 
matching approach. Further emissions can be generated by replacing $O(\Phi_1)$ with the 
parton-shower generating functional $\mc{F}_1(t_1,O)$, where
\begin{equation}
  \mc{F}_n(t,O)=\Pi_n(t_c,t)\,O(\Phi_n)
  +\int_{t_c}^t\done\hat{\Phi}_1\,\mr{K}_n(\hat{\Phi}_1)\,
  \Pi_n(\hat{t},t)\,\mc{F}_{n+1}(\hat{t},O)\;,
  \quad\text{where}\quad
  \hat{t}=t(\hat{\Phi}_1)\;.
\end{equation}
We now replace the Born differential cross section in
Eq.~\eqref{eq:unlops_step1} by the differential NLO cross section
\begin{equation}\label{eq:unlops_inclusive_xs}
  \bar{\mr{B}}(\Phi_0)=\mr{B}_0(\Phi_0)+\tilde{\mr{V}}_0(\Phi_0)+\mr{I}_0(\Phi_0)
  +\int\done\hat{\Phi}_1\,\Big[\mr{B}_1(\Phi_0,\hat{\Phi}_1)
    -\mr{S}_0(\Phi_0,\hat{\Phi}_1)\Big]\;.
\end{equation}
$\tilde{\mr{V}}_0$ denotes the UV finite part of the virtual corrections, including 
collinear mass factorization counterterms, $\mr{I}_0$ are integrated NLO subtraction 
terms~\cite{Catani:1996vz,*Catani:2002hc}, and $\mr{S}_0$ the corresponding 
real subtraction terms. The matched result is given by
\begin{equation}\label{eq:unlops_step2}
  \begin{split}
    &\bigg\{\int\done\Phi_0\,\bar{\mr{B}}_0^{t_c}(\Phi_0)
    +\int_{t_c}\done\Phi_1\,\Big[1-\Pi_0(t_1,\mu_Q^2)\Big]\mr{B}_1(\Phi_1)\bigg\}\,O(\Phi_0)
    +\int_{t_c}\done\Phi_1\,\Pi_0(t_1,\mu_Q^2)\,\mr{B}_1(\Phi_1)\,\mc{F}_1(t_1,O)\;,
  \end{split}
\end{equation}
where we have defined the vetoed cross section
\begin{equation}\label{eq:unlops_vetoed_xs}
  \bar{\mr{B}}_0^{t_c}(\Phi_0) = \bar{\mr{B}}_0(\Phi_0)
  -\int_{t_c}\done\Phi_1\,\mr{B}_1(\Phi_1)\;.
\end{equation}
Equation~\eqref{eq:unlops_step2} already contains the essence of our method. 
The three terms can be generated in a Monte Carlo simulation as follows: 
$\bar{\mr{B}}_0^{t_c}$ is a fixed-order contribution, which does not undergo 
parton showering. $\mr{B}_1$ is assigned a parton shower ``history'' using the 
clustering procedure first proposed in~\cite{Andre:1997vh}. The zero-parton state
defined in this clustering undergoes truncated parton shower evolution.
By definition, the survival probability is $\Pi_0(t_1,\mu_Q^2)$,
while the corresponding branching probability is $1-\Pi_0(t_1,\mu_Q^2)$.
Thus, if an emission is generated, the event is kept in the selected zero-parton state,
as indicated by the observable dependence $O(\Phi_0)$ in the first term of Eq.~\eqref{eq:unlops_step2}.
If no emission is generated, the event undergoes parton showering, starting 
from the one-parton state. This procedure is an improvement of \UNLOPS and 
ensures that no counter-events with negative weights must be generated during
the matching.

Up to now we have ignored renormalization and factorization scale dependence in $B_1$.
While terms generated by the running of the strong coupling are formally of higher order
and therefore do not modify the fixed-order accuracy of the matched result,
they are important to restore the logarithmic accuracy of the parton shower.
The same reasoning applies to scaling violations in the PDFs.
Scales can be adjusted to their parton shower values by reweighting,
eventually leading to the improved \UNLOPS matching formula
\begin{equation}\label{eq:unlops}
  \begin{split}
    \abr{O}^{\rm(UNLOPS)}=&\bigg\{\int\done\Phi_0\,\bar{\mr{B}}_0^{t_c}(\Phi_0)
    +\int_{t_c}\done\Phi_1\,\Big[1-\Pi_0(t_1,\mu_Q^2)\,w_1(\Phi_1)\Big]\mr{B}_1(\Phi_1)\bigg\}\,O(\Phi_0)\\
    &+\int_{t_c}\done\Phi_1\,\Pi_0(t_1,\mu_Q^2)\,w_1(\Phi_1)\,\mr{B}_1(\Phi_1)\,\mc{F}_1(t_1,O)\;.
  \end{split}
\end{equation}
In the case of Drell-Yan lepton pair production we need to match a single initial-state 
parton splitting $a\to\{a',j\}$. The weight $w_1(\Phi_1)$ is then defined 
as~\cite{Lonnblad:2012ix}
\begin{equation}\label{eq:unlops_weight}
  w_1(\Phi_1)=\frac{\alpha_s(b\,t_1)}{\alpha_s(\mu_R^2)}\,
  \frac{f_a(x_a,t_1)}{f_a(x_a,\mu_F^2)}
  \frac{f_{a'}(x_{a'},\mu_F^2)}{f_{a'}(x_{a'},t_1)}
  \qquad\text{where}\qquad
  \beta_0\ln\frac{1}{b}=\bigg(\frac{67}{18}-\frac{\pi^2}{6}\bigg)C_A-\frac{10}{9}\,T_R\,n_f\;.
\end{equation}
$f_a(x_a)$ and $f_{a'}(x_{a'})$ denote the PDFs associated with the external
and intermediate parton, respectively.
The scale factor $b$ includes effects of the 2-loop cusp anomalous dimension 
in the parton shower~\cite{Catani:1990rr}.

The event generation procedure is modified as follows:
Weights of 1-jet events are multiplied by $1+2|w_1-1|$. In a fraction 
$1/(2+1/|w_1-1|)$, the event is weighted by $\mr{sgn}(w_1-1)$, and the point 
is discarded if an emission is generated in the truncated parton shower.
If the event is kept, it is reduced to Born kinematics and the sign of its 
weight inverted with probability 1/2.
This procedure sums -- in a Monte-Carlo fashion -- over two event types,
which either contain factors of $\Pi_0$ or $1-\Pi_0$, or else
the terms $\pm\Pi_0(w_1-1)$.
This can lead again to the generation of negative weights, however their
fraction is much reduced compared to the original \UNLOPS scheme.

Equation~\eqref{eq:unlops} still holds if the phase-space separation is not achieved
in terms of the parton-shower evolution parameter, i.e.\ if the integration
boundaries for $\bar{\mr{B}}_0^{t_c}$ and $\int_{t_c}\done\Phi_1\,\mr{B}_1$ 
are not given by $t_c$.
In this case, one can split the real-emission contribution into a pure 
fixed-order part and a contribution where parton-shower resummation is applied.
In the following, we therefore define $\Pi_n(t,t')=\Pi_n(t_c,t')$ for all $t<t_c$.

We conclude this section with a comparison to the \POWHEG method~\cite{Nason:2004rx,*Frixione:2007vw}.
Assuming that the parton-shower evolution kernels for the first emission can be replaced by
$\mr{K}_0\to\bar{\mr{K}}_0=w_1\,\mr{B}_1/\mr{B}_0$, we obtain from Eq.~\eqref{eq:unlops}
\begin{equation}\label{eq:unlops_comp}
  \begin{split}
    &\abr{O}^{\rm(UNLOPS)}\to\int\done\Phi_0\,\mr{B}_0(\Phi_0)\,\bar{\mc{F}}_0(\mu_Q^2,O)
    +\int\done\Phi_0\,\Big[\bar{\mr{B}}_0(\Phi_0)-\mr{B}_0(\Phi_0)\Big]\,O(\Phi_0)\;.
  \end{split}  
\end{equation}
The main difference compared to the \POWHEG result,
\begin{equation}\label{eq:powheg_comp}
  \begin{split}
    \abr{O}^{\rm(POWHEG)}=&\int\done\Phi_0\,\bar{\mr{B}}_0(\Phi_0)\,\bar{\mc{F}}_0(\mu_Q^2,O)\;,
  \end{split}  
\end{equation}
is that the finite remainder of higher-order corrections (after UV renormalization
and IR subtraction), $\bar{\mr{B}}_0-\mr{B}_0$, does not undergo parton showering
in \UNLOPS, while it does in \POWHEG. A comparison with \MCatNLO 
leads to the same conclusion. While it is not obvious from the matching
conditions at NLO, whether \UNLOPS or \POWHEG is the more natural prescription,
the NNLO matching conditions require that \UNLOPS at NNLO behaves identical
to both \MCatNLO and \POWHEG in this regard, i.e.\ that the finite remainder
multiplies, $\mc{F}_0(\mu_Q^2,O)$, rather than $O(\Phi_0)$. We
will return to this question at the end of section \ref{sec:unmcatnlo}.

\section{\protect\UNNLOPS with phase-space slicing}
\label{sec:unmcatnlo}
We first describe our calculation of the NNLO vetoed
cross section, corresponding to Eq.~\eqref{eq:unlops_vetoed_xs}.
It is performed in the $q_T$ subtraction method~\cite{Catani:2007vq,*Catani:2009sm}
with a $q_T$ cutoff. All soft and collinear singularities of NNLO origin
cancel within the zero-$q_T$ bin, leading to a finite differential 
cross section, $\bar{\bar{\mr{B}}}_0^{q_{T,\mr{cut}}}$.
The remainder is computed as an NLO result for the original Born process
plus one additional jet.

The NNLO cross section with a small cut on observables
like $q_T$ of the gauge boson has a simple factorization formula, which
can be described up to power corrections in the cutoff, $q_{T,\rm cut}$,
by effective field theory. This form is generally more compact than the full NNLO result.
The cutoff method has been used previously to compute top decay fully exclusively
at NNLO~\cite{Gao:2012ja}. We adopt the framework developed in~\cite{Becher:2010tm}
to obtain the vetoed cross section. All components needed for two loop results 
for the Drell-Yan process have recently been computed~\cite{Gehrmann:2012ze,
  *Gehrmann:2014yya,*Gehrmann:2014uaa}, and verified against the hard collinear 
coefficients~\cite{Catani:2012qa} used by the original $q_T$ subtraction method.

The contribution at $q_T>q_{T,\rm cut}$ is computed as a standard NLO QCD 
result, using Catani-Seymour dipole subtraction to regularize infrared
divergences~\cite{Catani:1996vz,*Catani:2002hc}. This type of calculation
has been fully automated~\cite{Binoth:2010xt,*Alioli:2013nda}. We
use \Sherpa~\cite{Gleisberg:2003xi,*Gleisberg:2008ta,Gleisberg:2007md}
for tree-level like contributions and \BlackHat~\cite{Berger:2008sj,
  *Berger:2009ep,*Berger:2010vm,*Berger:2010zx} for virtual corrections.
We match this computation to the parton shower using a variant of the \MCatNLO
method, which is described in~\cite{Hoeche:2011fd,*Hoeche:2012ft,*Hoeche:2012fm}.
The corresponding expression for the $q_T>q_{T,\rm cut}$ cross section
depending on an infrared-safe observable $O$ is
\begin{equation}\label{eq:mcnlo_one_jet}
  \abr{O}^\mr{(NLO)}_{>q_{T,\mr{cut}}}=\int_{q_{T,\mr{cut}}}\bs
  \done\Phi_1\tilde{\mr{B}}_1(\Phi_1)\,\tilde{\mc{F}}_1(t_1,O)
  +\int_{q_{T,\mr{cut}}}\bs\done\Phi_2\mr{H}_1(\Phi_2)\,\mc{F}_2(t_2,O)\;,
\end{equation}
where the one-jet differential NLO cross section and hard remainder are defined as
\begin{equation}\label{eq:def_bbar_h_mcnlo}
  \begin{split}
  \tilde{\mr{B}}_1(\Phi_1)=&\;\mr{B}_1(\Phi_1)+\tilde{\mr{V}}_1(\Phi_1)+\mr{I}_1(\Phi_1)
  -\int_{t_c}\done\hat{\Phi}_1\,\mr{S}_1(\Phi_1,\hat{\Phi}_1)\,\Theta(t_2(\hat{\Phi}_1)-t_1(\Phi_1))\;,\\
  \mr{H}_1(\Phi_2)=&\;\mr{B}_2(\Phi_2)-\mr{S}_1(\Phi_2)\,\Theta(t_1(\Phi_2)-t_2(\Phi_2))\;.
  \end{split}
\end{equation}
The generating functional of the \MCatNLO is
\begin{equation}
  \tilde{\mc{F}}_1(t,O)=\tilde{\Pi}_1(t_c,t_1)\,O(\Phi_1)
      +\int_{t_c}\done\hat{\Phi}_1\frac{\mr{S}_1(\Phi_1,\hat{\Phi}_1)}{\mr{B}_1(\Phi_1)}\,
      \tilde{\Pi}_1(\hat{t},t_1)\,\mc{F}_2(\hat{t},O)\;,
\end{equation}
with the no-branching probability given by parton-shower unitarity:
\begin{align}
  \tilde{\Pi}_1(t,t',\Phi_n)=&\exp\left\{-\int_t^{t'}\done\hat{\Phi}_1\,
  \frac{\mr{S}_1(\Phi_1,\hat{\Phi}_1)}{\mr{B}_1(\Phi_1)}\right\}\;.
\end{align}
Note that we choose $q_{T,\mr{cut}}\leq 1~\mr{GeV}$,
which is below the cutoff of the initial-state parton shower.

Equation~\eqref{eq:mcnlo_one_jet} produces the correct dependence on the
observable $O$ at next-to-leading QCD for $q_T>q_{T,\mr{cut}}$. 
It can thus be used to complement the exclusive NNLO calculation in the zero-$q_T$ bin.
However, the two calculations cannot be naively added as in Eq.~\eqref{eq:unlops},
since this would spoil the $\mc{O}(\alpha_s^2)$ accuracy of the full result.
This problem was also addressed by NLO merging methods~\cite{
  Lavesson:2008ah,Gehrmann:2012yg,*Hoeche:2012yf,Lonnblad:2012ix,Platzer:2012bs},
and by the \MINLO scale setting procedure~\cite{Hamilton:2012rf}.
The $\mc{O}(\alpha_s)$ contribution to the fixed-order expansion of the parton shower
must first be subtracted, which can be achieved efficiently by omitting the first 
emission in a truncated shower~\cite{Gehrmann:2012yg,*Hoeche:2012yf}, or by explicit 
subtraction~\cite{Lavesson:2008ah,Lonnblad:2012ix}.
Correspondingly, any $\mc{O}(\alpha_s)$ contribution must be subtracted from
the corrective weight, Eq.~\eqref{eq:unlops_weight}.
The full formula describing our combination method can be written as
\begin{equation}\label{eq:nnlo_ps}
  \begin{split}
    &\abr{O}^\mr{(UN^2LOPS)}=\;
    \int\done\Phi_0\,\bar{\bar{\mr{B}}}_0^{q_{T,\mr{cut}}}(\Phi_0)\,O(\Phi_0)\\
    &\quad+\int_{q_{T,\mr{cut}}}\bs\done\Phi_1\,
    \Big[1-\Pi_0(t_1,\mu_Q^2)\,
      \Big(w_1(\Phi_1)+w_1^{(1)}(\Phi_1)+\Pi_0^\mr{(1)}(t_1,\mu_Q^2)\Big)\Big]\,
    \mr{B}_1(\Phi_1)\,O(\Phi_0)\\
    &\quad+\int_{q_{T,\mr{cut}}}\bs\done\Phi_1\,
    \Pi_0(t_1,\mu_Q^2)\Big(w_1(\Phi_1)+w_1^{(1)}(\Phi_1)+\Pi_0^\mr{(1)}(t_1,\mu_Q^2)\Big)
    \,\mr{B}_1(\Phi_1)\,\bar{\mc{F}}_1(t_1,O)\\
    &\quad+\int_{q_{T,\mr{cut}}}\bs\done\Phi_1\,
      \Big[1-\Pi_0(t_1,\mu_Q^2)\Big]\,\tilde{\mr{B}}_1^{\rm{R}}(\Phi_1)\,O(\Phi_0)
    +\int_{q_{T,\mr{cut}}}\bs\done\Phi_1
    \Pi_0(t_1,\mu_Q^2)\,\tilde{\mr{B}}_1^{\rm{R}}(\Phi_1)\,\bar{\mc{F}}_1(t_1,O)\\
    &\quad+\int_{q_{T,\mr{cut}}}\bs\done\Phi_2\,
      \Big[1-\Pi_0(t_1,\mu_Q^2)\Big]\,\mr{H}_1^{\mr{R}}(\Phi_2)\,O(\Phi_0)
    +\int_{q_{T,\mr{cut}}}\bs\done\Phi_2\,
      \Pi_0(t_1,\mu_Q^2)\,\mr{H}_1^{\mr{R}}(\Phi_2)\,\mc{F}_2(t_2,O)\\
    &\quad+\int_{q_{T,\mr{cut}}}\bs\done\Phi_2\,
      \,\mr{H}_1^{\mr{E}}(\Phi_2)\,\mc{F}_2(t_2,O)
  \end{split}
\end{equation}
We have defined $\tilde{\mr{B}}^{\rm{R}}=\tilde{\mr{B}}-\mr{B}$ and the
regular and exceptional part of the hard remainder
\begin{align}\label{eq:unordered_hevents}
  \mr{H}_1^{\mr{R}}(\Phi_2)
  &= \mr{H}_1(\Phi_2)
     \Theta\left(t_1-t_2\right)
     \Theta\left(t_2-t_c\right)\;,
  &\mr{H}_1^{\mr{E}}(\Phi_2)
  &= \mr{H}_1(\Phi_2)-\mr{H}_1^{\mr{R}}(\Phi_2)\;.
\end{align}
The exceptional contributions $\mr{H}_1^{\mr{E}}$ contain phase space regions for
which no ordered parton shower history can be identified, as well as two-parton states
that do not allow an interpretation as having evolved from a zero- or one-parton state
via QCD-type parton splittings. Exceptional contributions do not undergo the
truncated parton showering used to produce $\Pi_0(t_1,\mu_Q^2)$, as they do not generate
logarithmic corrections at parton shower accuracy. 
Ambiguities in the matched result due to exceptional configurations 
will be important for matching at higher logarithmic accuracy, and
can be resolved as soon as the parton shower is amended with the necessary sub-leading
logarithmic corrections and electroweak splittings. This will allow to treat such states
in the same manner as the regular configurations.

The subtraction terms for the no-branching probability of the parton shower,
and for the weight $w_1$, are given by 
\begin{equation}\label{eq:unnlops_weight_subterms}
  \begin{split}
    \Pi_0^{(1)}(t,t')=&\;\int_t^{t'}\done\hat{\Phi}_1\,
    \frac{\alpha_s(\mu_R^2)}{\alpha_s(b\,\hat{t})}\,\mr{K}_1(\Phi_1,\hat{\Phi}_1)\\
    w_1^{(1)}(\Phi_1)=&\;\frac{\alpha_s(\mu_R^2)}{2\pi}\Bigg[\,\beta_0\log\frac{b\,t_1}{\mu_R^2}
    -\log\frac{t_1}{\mu_F^2}\sum_c\bigg(\int_x^1\frac{\done z}{z}P_{ca}(z)\,
    \frac{f_c(x/z,\mu_F^2)}{f_a(x,\mu_F^2)}
    -\int_{x'}^1\frac{\done z}{z}P_{c{a'}}(z)\,
    \frac{f_c(x'/z,\mu_F^2)}{f_{a'}(x',\mu_F^2)}\bigg)\Bigg]\;.
  \end{split}
\end{equation}
They are generated by the Monte-Carlo procedure outlined below Eq.~\eqref{eq:unlops_weight}.
Note that $1-\Pi_0\big(w_1+w_1^{(1)}+\Pi_0^{(1)}\big)$ is of order $\alpha_s^2$. 
Therefore, it is easy to see that the method does not spoil the accuracy
of the fixed-order calculation. To investigate if the logarithmic accuracy
of the parton shower resummation is maintained, we take the collinear limit,
$t_1\to 0$. In this limit,  $\mr{H}_1^{\mr{E}}$ only generates
logarithms that are beyond the parton shower approximation, and it can thus 
be ignored. Consequently, for $q_T>q_{T,\rm cut}$, we are left with
\begin{equation}\label{eq:unnlops_1j_bin}
  \int_{q_{T,\mr{cut}}}\bs\done\Phi_1\,
  \Pi_0(t_1,\mu_Q^2)\,w_1(\Phi_1)\,\mr{B}_1(\Phi_1)\,\bar{\mc{F}}_1(t_1,O)
  +\int_{q_{T,\mr{cut}}}\bs\done\Phi_1\,
  \Pi_0(t_1,\mu_Q^2)\,\mr{R}_1(\Phi_1,O)
\end{equation} 
where
\begin{equation}\label{eq:unnlops_remainder}
  \begin{split}
  \mr{R}_1(\Phi_1,O)=&\,\bigg(\mr{B}_1(\Phi_1)\,
  \Big(w_1^{(1)}(\Phi_1)+\Pi_0^\mr{(1)}(t_1,\mu_Q^2)\Big)
    +\tilde{\mr{B}}_1^{\rm{R}}(\Phi_1)\bigg)\,\bar{\mc{F}}_1(t_1,O)
      +\int\done\hat{\Phi}_1\,\mr{H}_1^{\mr{R}}(\Phi_1,\hat{\Phi}_1)\,\mc{F}_2(t_2,O)
  \end{split}
\end{equation} 
The first term in Eq.~\eqref{eq:unnlops_1j_bin} is, to the required accuracy,
equivalent to the parton shower result. Thus it remains to be shown that
$\mr{R}_1$ contains only subleading terms.
In the soft and collinear limit, $\mr{H}_1^{\mr{R}}$ does not contribute 
at the required accuracy~\cite{Frixione:2002ik}. Quark propagators in $\tilde{\mr{B}}_1^{\rm{R}}$ 
can to first order be approximated as $(1-\Pi_0^{(1)}(t,\mu_Q^2)-w_1^{(1)}
  +\alpha_s/(2\pi)\beta_0\log(b\,t/\mu_R^2))/p\!\!\!\slash$~\cite{Amati:1980ch}, 
where $p$ is the quark momentum. Coupling renormalization leads to corrections 
of the form $\alpha_s/(2\pi)\beta_0\log(t/\mu_R^2)$, where $t$ is the
relative transverse momentum in the gluon emission~\cite{Dokshitzer:1978qu,
  *Amati:1978by,*Ellis:1978sf,*Libby:1978ig,*Mueller:1978xu,*Dokshitzer:1978hw}.
The two-loop cusp anomalous dimension, simulated by means of the scale factor 
$b$ in Eq.~\eqref{eq:unlops_weight}, is naturally present in $\tilde{\mr{B}}_1^{\rm{R}}$.
The subtraction terms, Eq.~\eqref{eq:unnlops_weight_subterms}, thus cancel
all universal NLO corrections in $\tilde{\mr{B}}_1^{\rm{R}}$, which have
already been included in the parton shower. The remainder is beyond the
required accuracy. Using the unitarity condition for parton shower evolution,
this argument extends to the region $q_T<q_{T,\rm cut}$. Note that because 
of the unitarity condition, also no spurious logarithms are generated 
in the inclusive cross section, and the NNLO accuracy is maintained exactly.

We now return to the difference between \UNLOPS and \POWHEG/\MCatNLO discussed in 
Sec.~\ref{sec:intro}. When performing a one-jet matched NLO calculation
in the \UNLOPS implementation of~\cite{Lonnblad:2012ix}, the non-universal terms 
in the first part of Eq.~\eqref{eq:unnlops_remainder} do not undergo parton showering
above the merging scale. The \UNNLOPS prescription instead introduces parton shower 
corrections to these terms throughout the resummation phase space, and it includes 
a Sudakov form factor for a truncated shower to resum effects of unresolved emissions 
above the scale of the hard jet. This new matching condition is justified
if we view the parton shower as an all-order calculation dressing a hard input
state, which has a fixed-order expansion by itself, with the effects of soft and
collinear radiation\footnote{A similar interpretation holds for factorization
formulae in analytic resummation, for which a fixed-order hard function is
convoluted with all-order soft and collinear functions, see 
for example~\cite{Becher:2010tm} and~\cite{Catani:2013tia}.}.
It can also be understood in the following way: In the collinear limit, 
the factorization of one-loop matrix elements leads to virtual corrections 
of the form $\mr{V}_0\mr{K}_0+\mr{B}_0\mr{K}_0^{\textnormal{\tiny(1)}}$, 
where $\mr{K}_0^{\textnormal{\tiny(1)}}$ are the one-loop splitting kernels.
When including the respective integrated subtraction terms and no-branching
probabilities of the truncated parton shower, the remainder of the first term
turns into $(\bar{\mr{B}}_0-\mr{B}_0)\mr{K}_0\,\Pi_0$, and can be interpreted
as a parton shower combined with the finite remainder of the NLO corrections
in the zero-$q_T$ bin. This eliminates the difference between 
Eqs.~\eqref{eq:powheg_comp} and~\eqref{eq:unlops_comp}.

Following this argument, one may conclude that the finite $\mc{O}(\alpha_s^2)$ 
corrections contained in zero-$q_T$ configurations of Eq.~\eqref{eq:nnlo_ps} should 
also be ``spread'' across the one-parton (and two-parton) phase space by the parton shower, 
provided that the resulting change of the radiation pattern is at most $\mathcal{O}(\alpha_s^3)$.
The difference between including and not including such parton shower corrections
is within the intrinsic uncertainty of NNLO matching schemes.
We see no strong reason for implementing them in the simulation of the Drell-Yan
process, due to the excellent convergence of the perturbative series. The
assessment may differ in other reactions, like Higgs-boson production via gluon
fusion, where higher-order corrections are large.

\section{Results}
\label{sec:results}
This section presents results using an implementation of the
\UNNLOPS algorithm in the event generator \Sherpa~\cite{Gleisberg:2003xi,*Gleisberg:2008ta}, 
We use a parton shower~\cite{Schumann:2007mg} based on Catani-Seymour
dipole subtraction~\cite{Catani:1996vz,*Catani:2002hc}. NLO virtual corrections for the
one-jet process are provided by the \BlackHat library~\cite{Berger:2008sj,
  *Berger:2009ep,*Berger:2010vm,*Berger:2010zx}. Dipole subtraction is performed using
Amegic~\cite{Krauss:2001iv,Gleisberg:2007md}. For comparison to experimental data
we use Rivet~\cite{Buckley:2008vh,*Buckley:2010ar}. We use the MSTW 2008 
PDF set~\cite{Martin:2009iq} and the corresponding definition of the 
running coupling. We work in the five flavor scheme. Electroweak parameters are given
in the $G_\mu$ scheme as $m_Z=91.1876\;{\rm GeV}$, $\Gamma_Z=2.4952\;{\rm GeV}$,
$m_W=80.385\;{\rm GeV}$, $\Gamma_W=2.085\;{\rm GeV}$
and $G_F=1.1663787\cdot 10^{-5}\;{\rm GeV}^{-2}$.

\begin{figure}
  \includegraphics[width=0.475\textwidth]{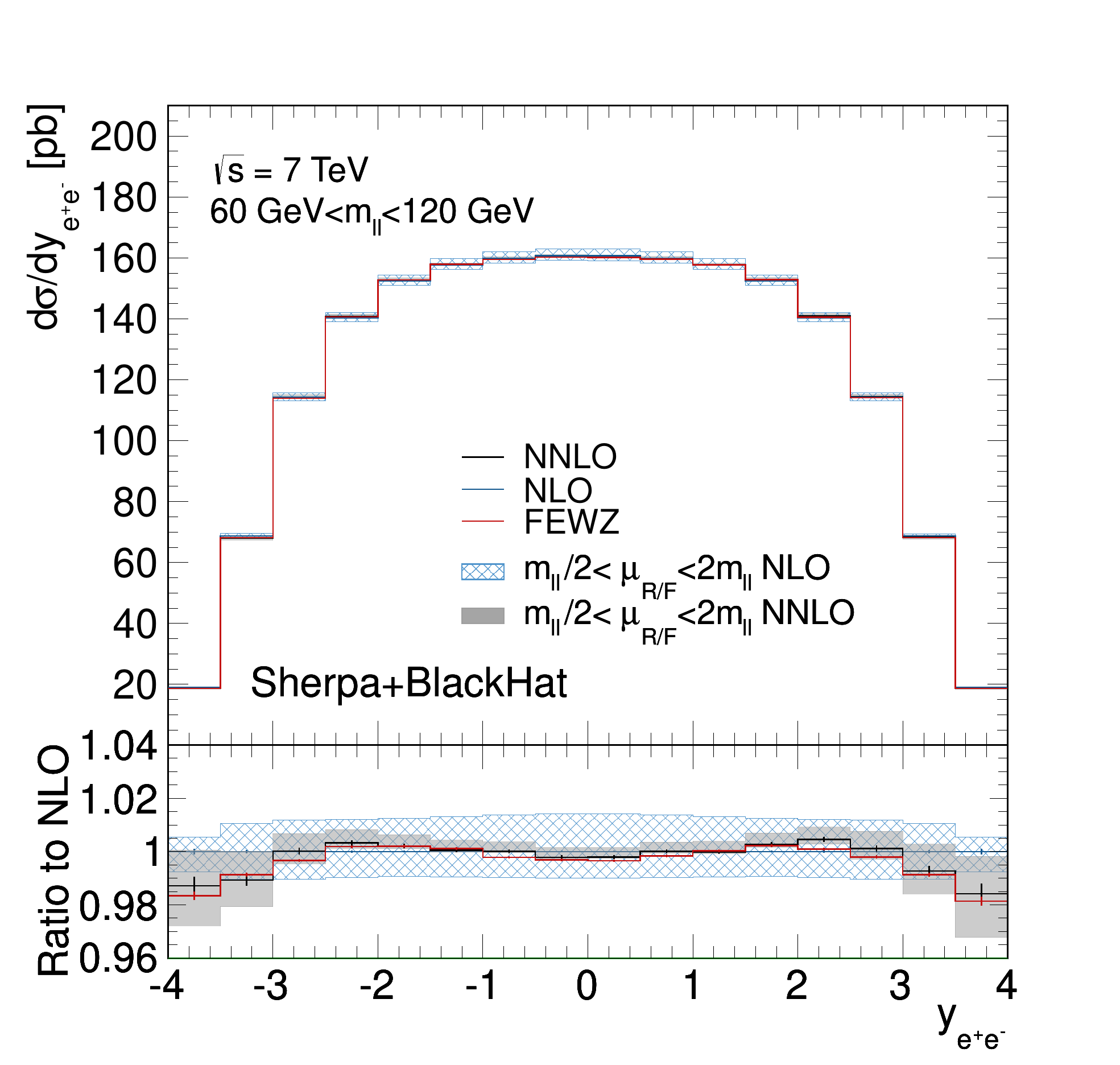}
  \includegraphics[width=0.475\textwidth]{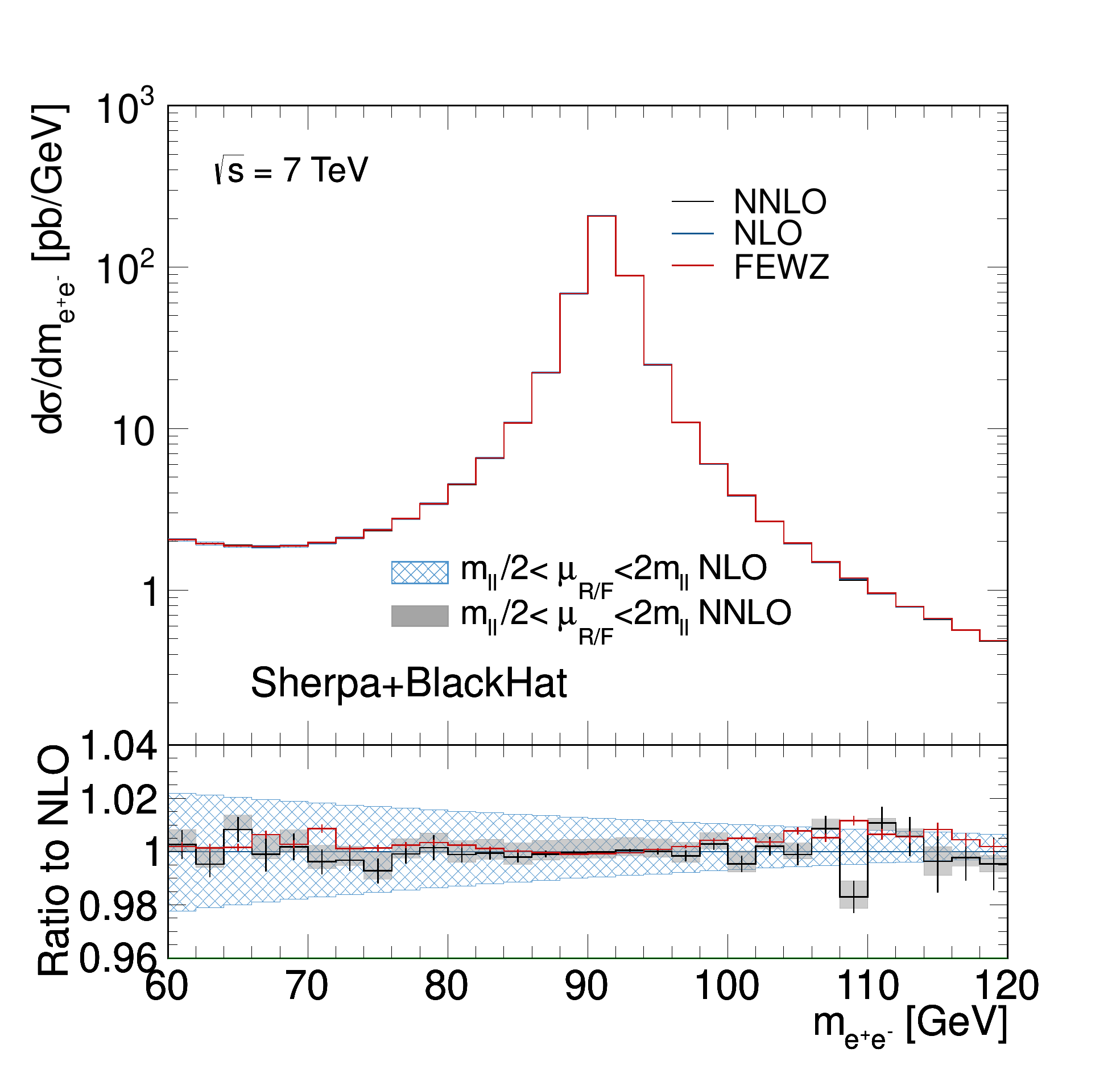}
  \caption{Comparison between \FEWZ and \Sherpa for rapidity
    and invariant mass spectra of the Drell-Yan lepton pair.
    The gray solid (brown hatched) band shows scale uncertainties associated
    with the NNLO (NLO) prediction, obtained by varying
    $\mu_{R/F}$ in the range $m_{ll}/2\le\mu\le 2\,m_{ll}$.
    \label{fig:fo_y_m}}
\end{figure}

In order to cross-check our implementation we first compare the total
cross section in the mass range $60\;{\rm GeV}\le m_{ll}\le 120\;{\rm GeV}$
to results obtained from VRAP~\cite{Anastasiou:2003yy,*Anastasiou:2003ds}.
Table~\ref{tab:xs_comparison} shows that the predictions
agree within the permille-level statistical uncertainty of the 
Monte-Carlo integration. We also compared the central values to
results from \DYNNLO~\cite{Catani:2009sm} and found full agreement.
Additionally, we have checked that our predictions are identical
when varying $q_{T,\rm cut}$ between 0.1~GeV and 1~GeV.
The default value is $q_{T,\rm cut}=$1~GeV.
\begin{table}
  \begin{tabular}{l@{\hspace*{5mm}}r@{}l@{}l@{\hspace*{5mm}}r@{}l@{}l
      @{\hspace*{5mm}}r@{}l@{}l@{\hspace*{5mm}}r@{}l@{}l}\hline
    $E_{\rm cms}$ & \multicolumn{3}{c}{7 TeV} & \multicolumn{3}{c}{14 TeV} &
    \multicolumn{3}{c}{33 TeV} & \multicolumn{3}{c}{100 TeV} \\\hline\hline
    VRAP \vphantom{$\int^{^A}_{_B}$} &
    973.&99(9)&$^{+4.70}_{-1.84}$ pb&
    2079.&0(3)&$^{+14.7}_{-6.9}$ pb&
    4909.&7(8)&$^{+45.1}_{-27.2}$ pb&
    13346&(3)&$^{+129}_{-111}$ pb\\
    SHERPA &
    973.&7(3)&$^{+4.78}_{-2.21}$ pb&
    2078.&2(10)&$^{+15.0}_{-8.0}$ pb&
    4905.&9(28)&$^{+45.1}_{-27.9}$ pb&
    13340&(14)&$^{+152}_{-110}$ pb\\[1mm]
    \hline
  \end{tabular}
  \caption{Total cross sections for $60\;{\rm GeV}\le m_{ll}\le 120\;{\rm GeV}$
    at varying center-of-mass energy for a $pp$-collider. Uncertainties from 
    scale variations are given as sub-/superscripts. Statistical uncertainties 
    from Monte-Carlo integration are quoted in parentheses.
    \label{tab:xs_comparison}}
\end{table}
Figure~\ref{fig:fo_y_m} compares differential cross sections from
\FEWZ~\cite{Gavin:2010az,*Gavin:2012sy,*Li:2012wna}
and \Sherpa for the rapidity and invariant mass spectra of the Drell-Yan 
lepton pair. It is interesting to observe the excellent agreement between
the NLO and NNLO predictions.

\begin{figure}
  \includegraphics[width=0.475\textwidth]{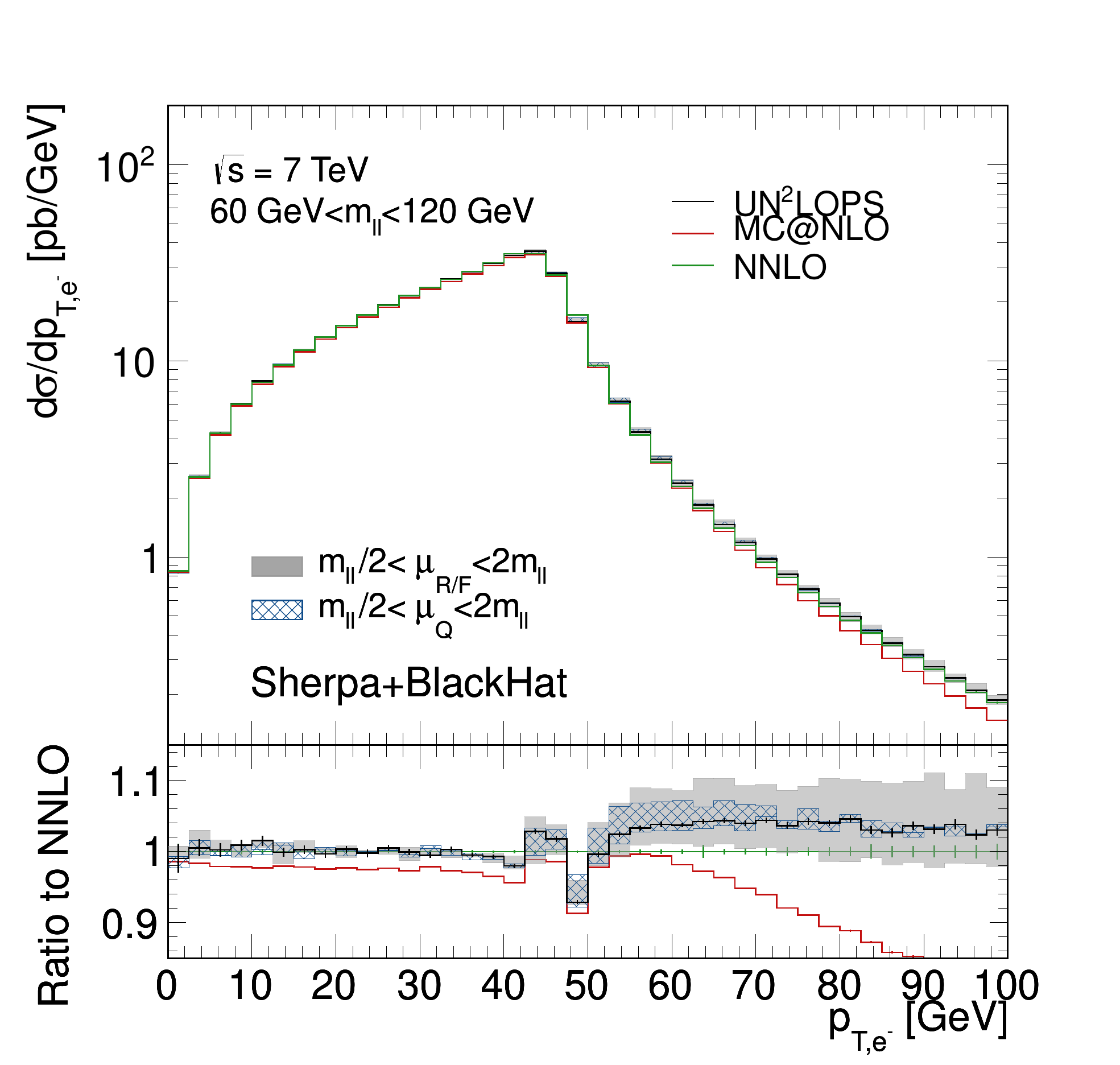}
  \includegraphics[width=0.475\textwidth]{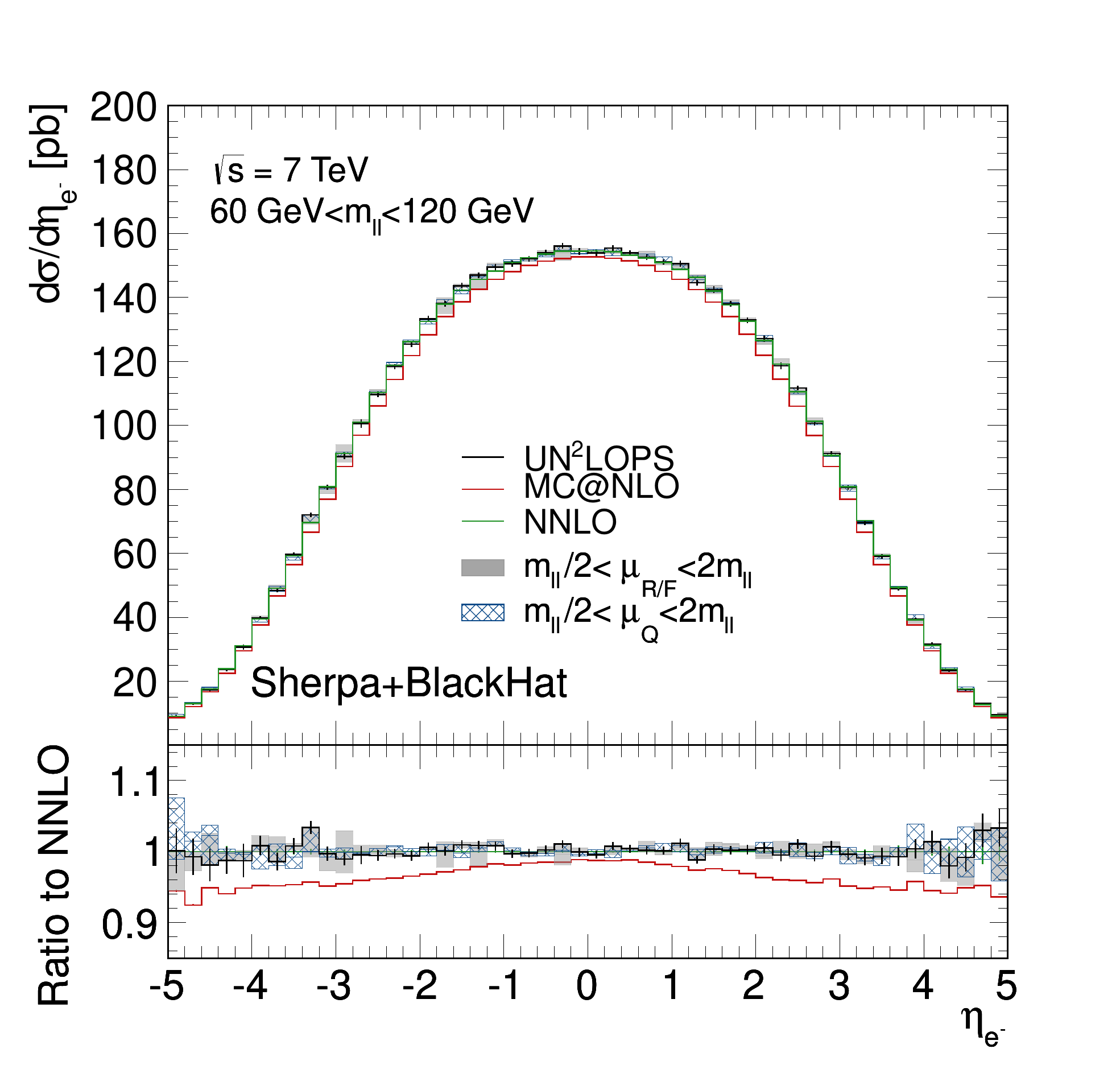}
  \caption{Transverse momentum and rapidity spectrum of the electron.
    The gray solid (blue hatched) band shows scale uncertainties obtained 
    by varying $\mu_{R/F}$ ($\mu_Q$) in the range
    $m_{ll}/2\le\mu\le 2\,m_{ll}$.
    \label{fig:ps_pte_etae}}
\end{figure}

Figure~\ref{fig:ps_pte_etae} shows predictions from the matched calculation.
We now include a simulation of higher-order QED corrections~\cite{Schonherr:2008av}.
It is interesting to compare the matched prediction to the
fixed-order NNLO result for the transverse momentum spectrum of the electron.
In the region $p_{T,e}<45~\mr{GeV}$ the result is generically NNLO correct,
while for $p_{T,e}>45~\mr{GeV}$, it is effectively only NLO correct. 
Correspondingly, the uncertainty band is larger at high transverse momentum.
The fixed-order prediction lies well within the NNLO scale uncertainty of the 
matched result, except for the transition region $p_{T,e}${\smaller$\gtrsim$}$45~\mr{GeV}$,
where real emission corrections play the dominant role.

\begin{figure}
  \includegraphics[width=0.475\textwidth]{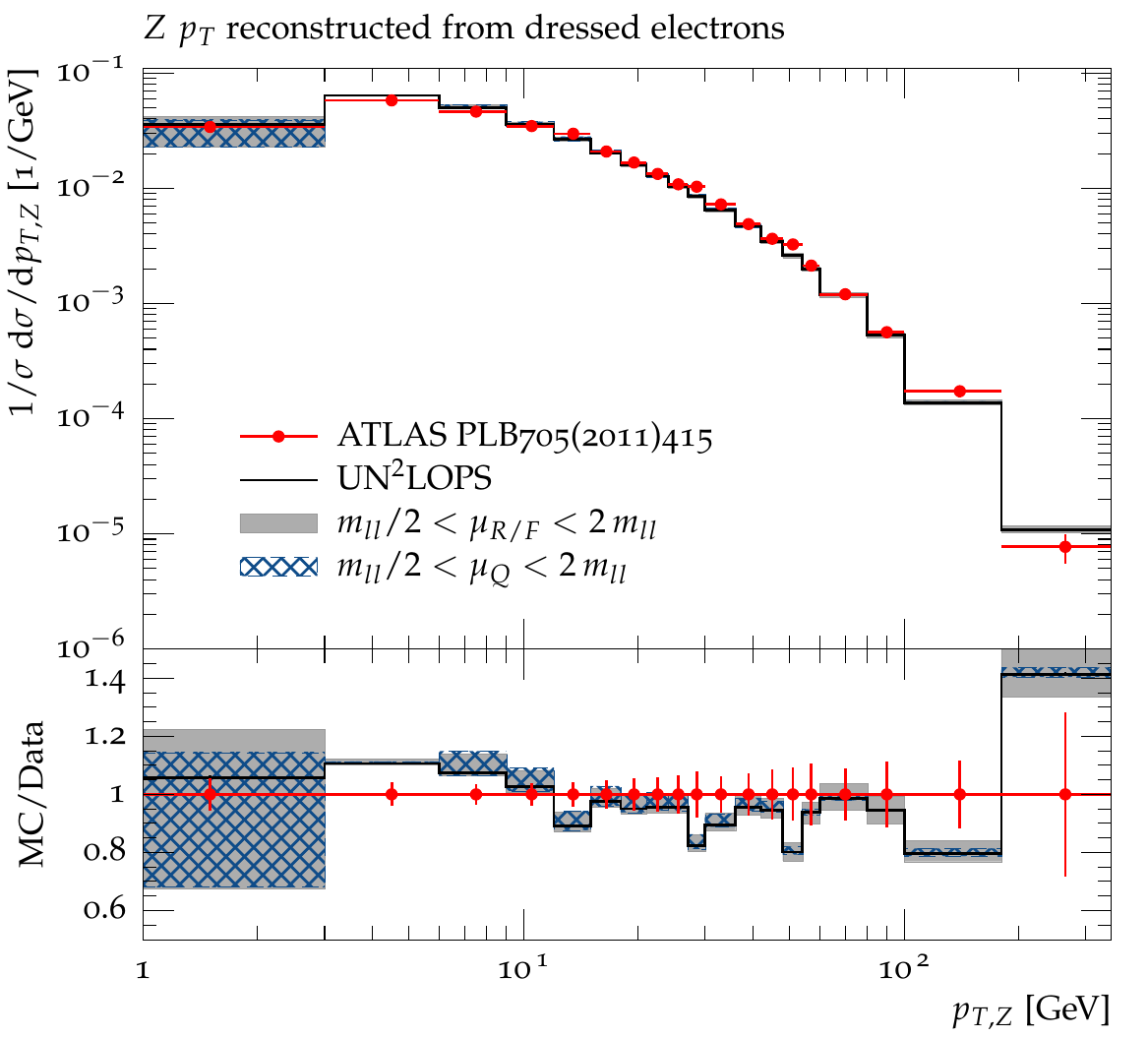}
  \includegraphics[width=0.475\textwidth]{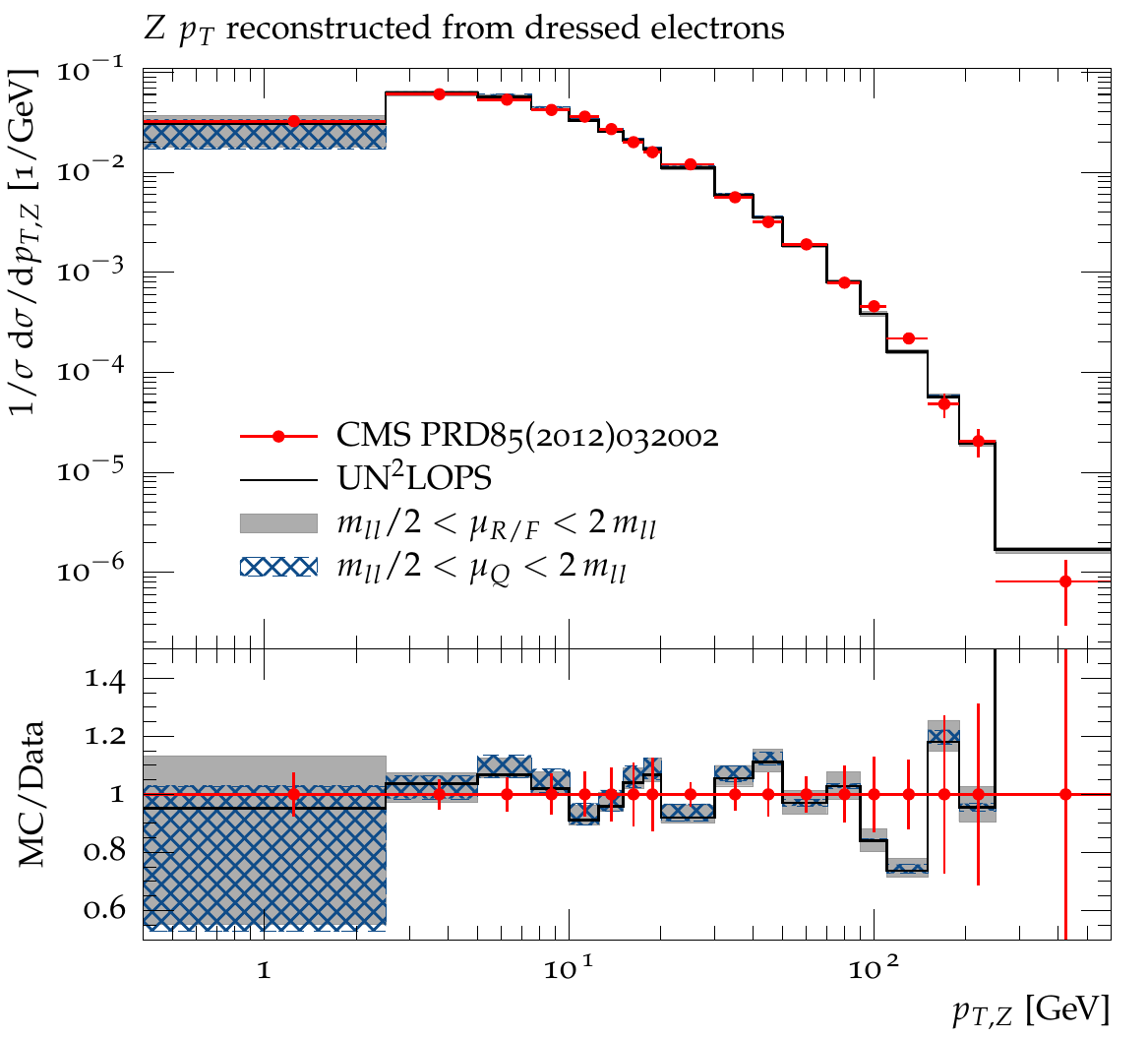}
  \caption{\UNNLOPS prediction for the transverse momentum spectrum
    of the Drell-Yan lepton pair in comparison to
    ATLAS data from~\cite{Aad:2011gj} (left)
    and CMS data from~\cite{Chatrchyan:2011wt} (right).
    The gray solid (blue hatched) band shows scale uncertainties 
    obtained by varying $\mu_{R/F}$ ($\mu_Q$) in the range
    $m_{ll}/2\le\mu\le 2\,m_{ll}$.
    \label{fig:ps_y_phis}}
\end{figure}

Figure~\ref{fig:ps_y_phis} compares the transverse momentum spectrum 
of the Drell-Yan lepton pair to data from the CMS~\cite{Chatrchyan:2011wt}
and ATLAS collaboration~\cite{Aad:2011gj}. 
These measurements are insensitive to generic NNLO corrections, 
which are generated only in the zero-$q_T$ bin in our approach.
However, they probe the form of the Sudakov form factor as simulated by the
matched calculation, and they are therefore useful to judge whether the
radiation pattern of the parton shower is preserved. The results indicate 
that higher-logarithmic corrections originating in $\tilde{\mr{B}}_1^{\rm R}$
and $\mr{H}_1$ are numerically small and do not spoil our prediction. Note that 
the parton-shower parameters in the matched calculation have not been tuned
to fit either of these distributions. The large perturbative uncertainties 
in the first bin of both distributions do not lead to large uncertainties 
in the total cross section, but they indicate that higher-logarithmic
resummation might be needed in order to improve the low-$p_{T,Z}$ region.

\section{Outlook}
\label{sec:conclusions}
We have presented a simple method for matching NNLO calculations in
perturbative QCD to existing parton showers, based on the \UNLOPS technique.
In contrast to the original implementation of \UNLOPS, the event generation
algorithm does not lead to large cancellations, and convergence 
of the Monte Carlo integration is much improved. 
Remaining uncertainties of the method are related to the treatment of
finite remainders of the virtual corrections after UV renormalization
and IR subtraction, and to the treatment of exceptional
configurations in the hard remainder of double real corrections.
Our method can be applied to arbitrary processes, and it can be 
systematically improved by using parton showers with higher 
logarithmic accuracy, which is currently an area of active research.
The combination with higher-multiplicity NLO matched simulations 
is straightforward and can be achieved in both the \UNLOPS~\cite{Lonnblad:2012ix}
and \MEPSatNLO~\cite{Gehrmann:2012yg,*Hoeche:2012yf} schemes.

We also provide an independent implementation of a fully differential
NNLO calculation of Drell-Yan lepton pair production, using the $q_T$-cutoff
method. Both the parton-level event generator and the shower-matched
calculation are made publicly available in the framework of the \Sherpa
event generator. This also allows the production of LHEF files~\cite{
  Alwall:2006yp,*Butterworth:2014efa} or NTuple files~\cite{Brun:1997pa,
  *Bern:2013zja} containing NNLO event information at parton level.
\smallskip

\begin{acknowledgments}
We are grateful to the \BlackHat collaboration for making the \BlackHat 
library available for this study. We thank Lance Dixon, Frank Krauss, 
Leif L{\"o}nnblad and HuaXing Zhu for helpful discussions and comments 
on the manuscript. This work was supported by the US Department of Energy 
under contract DE--AC02--76SF00515. We used resources of the National Energy Research 
Scientific Computing Center, which is supported by the Office of Science 
of the U.S.\ Department of Energy under Contract No.\ DE--AC02--05CH11231.
\end{acknowledgments}

\bibliography{journal}
\end{document}